# Influence of Defects on Barrier Energy Formation for OOH$^*$ Intermediate in ORR on Tetragonal-ZrO$_2$ with Adsorbed-Hydroxyl

Sara Fazeli$^{1*}$, Pascal Brault$^{1*}$, Amaël Caillard$^1$, Anne-Lise Thomann$^1$, Eric Millon$^1$,

Christophe Coutanceau$^2$, Soumya Atmane$^1$

$^1$GREMI, CNRS - Université d'Orléans, 45067 ORLEANS Cedex 2, France
$^2$ IC2MP, Université de Poitiers / CNRS, 4 rue Michel Brunet, 86022 Poitiers, France

**Abstract**

Accelerating the oxygen reduction reaction (ORR) is a main subject of electrocatalysis research. A critical step of ORR is the formation of the hydroperoxyl functional group (OOH$^*$) intermediate. In this study, we investigate the influence of defects on facilitating the creation of OOH$^*$ in a zirconia-based cathode under hydroxyl group (-OH) adsorption. Simulations involve tetragonal pristine ZrO$_2$ (111) surfaces with introduced oxygen vacancy (t-ZrO$_{2-x}$) and nitrogen dopant (ZrO$_{2-x}$N$_x$). Density functional theory (DFT) is used to calculate the competitive -OH adsorption energies on pristine and defective surfaces. It reveals that oxynitride t-ZrO$_{2-x}$N$_x$ and under-stoichiometric oxide t-ZrO$_{2-x}$ exhibit the lowest and highest susceptibility to -OH adsorption, respectively. Additionally, we have determined the Minimum Energy Pathway (MEP) for OOH$^*$ formation on t-ZrO$_2$, t-ZrO$_{2-x}$, and t-ZrO$_{2-x}$N$_x$ with adsorbed-OH using the Nudged Elastic Band (NEB) approach with the COMB3 potential. Our results highlight the significant influence of defects on tuning the barrier energy of OOH* formation. The trend in the barrier energy formation of OOH$^*$ decreases in the order of t-ZrO$_{2-x}$ > pristine t-ZrO$_2$ > t-ZrO$_{2-x}$N$_x$. We demonstrate that ZrO$_{2-x}$N$_x$ is a promising candidate for accelerating ORR due to its lower barrier energy for OOH$^*$ creation. The findings from this study offer crucial insights for experimentalists aiming to develop optimal non-platinum-based cathode materials.

**Keywords:** Oxygen reduction reaction, NEB, under-stoichiometric zirconium oxide, zirconium oxynitride, barrier energy

Over the past few decades, proton exchange membrane fuel cells (PEMFCs)[1-4] have been known as promising candidates for energy conversion devices in future hydrogen-based applications. However, next-generation energy technologies, like regenerative cells, and metal-air batteries, depend on oxygen reduction reaction (ORR)[5-8] and hydrogen oxidation reactions(HOR)[9,10] kinetics. In fuel cells, HOR and ORR take place at the anode and cathode, respectively. It is worth noting that the ORR in PEMFCs operates over six orders of magnitude slower than HOR at the anode in aqueous solutions[11]. This delay stems from diverse pathways involving various O-containing intermediates (such as OOH*, O*, and OH*) affecting adsorption, desorption, and reactions[12].

Decades of research addressed slow ORR kinetics, particularly with Pt-based catalysts[5,13,14], seeking cost-effective options. Transition metal oxides (TMO)[8,15,16] hold promise as heterogeneous ORR catalysts in acidic and alkaline media[17], facilitating reactions between catalysts and reactants in various phases, such as gas molecules adsorbed on a solid metal surface to yield a product. Particularly, tetragonal zirconia (t-$ZrO_2$)[8,18], mainly on the (111)[8] plane, stands out among various metal oxides as a highly effective catalyst for ORR due to its remarkable high-temperature stability[19] and robustness[20].

The catalytic effectiveness of zirconia relies on its stability in the tetragonal phase, which is attainable through doping[21], induced oxygen vacancy($V_O$)[22,23], and specific preparation techniques. For instance, increasing oxygen vacancy ($V_O$) concentration or combining ultrafine $ZrO_2$ nanoparticles (NPs) with nitrogen doping experimentally raised the onset potential for the (ORR)[24]. In addition, in the previous work[25], the superior electronic properties of two defective t-$ZrO_2$ structures (t-$ZrO_{2-x}$ and t-$ZrO_{2-x}N_x$) for catalytic applications compared to their pristine counterpart were highlighted. Muhammady et al.[26] presented initial density functional theory (DFT) findings on the impact of $V_O$ and N doping on ORR catalytic activities in t-$ZrO_2$ (101). Their conclusion highlighted the minor role of defects in increasing ORR catalytic activity on both surfaces. Thus, further research is needed to speed up the targeted design and optimization of catalysts for accelerating ORR. For instance, beyond introducing defects, the hydroxylation of zirconia's surface modifies its chemistry, creating catalytic sites for (ORR)[27]. These sites affect key ORR steps, such as oxygen adsorption, reduction, and intermediate species formation, resulting in a boost or decrease in the catalytic activity.

Añez et al.[28] examined hydroxyl (-OH) coverages on the t-$ZrO_2$ (110) surface, revealing that hydroxylation significantly stabilizes the t-$ZrO_2$ (110) surface. Nevertheless, they did not explore the impact of -OH adsorption on the (ORR) mechanism. It is noteworthy that the hydroxylation of t-ZrO$_2$ surface plays a dual role, potentially as a poisoning factor and simultaneously as a contributor to the (ORR) mechanism, especially in the creation of the pivotal OOH*[29] intermediate. Clearly, the optimal (4e$^-$) ORR involves O$_2$ hydrogenation to produce OOH*, followed by its hydrogenation to yield H$_2$O and atomic O in steps (1) and (2) of the subsequent ORR mechanisms (eq 1-5 Left) [30]. (where * refers to a Zr surface site). Evidently, the electrocatalytic activity for 4e$^-$ oxygen reduction is governed by the stability of the associated intermediates, OOH* and OH*[31].

The optimal (4 e$^-$) ORR mechanism

$$O_2 + * \rightarrow O_2^* \quad (1)$$
$$O_2^* + (H^+ + e^-) \rightarrow OOH^* \quad (2)$$
$$OOH^* + (H^+ + e^-) \rightarrow O^* + H_2O \quad (3)$$
$$O^* + (H^+ + e^-) \rightarrow OH^* \quad (4)$$
$$OH^* + (H^+ + e^-) \rightarrow H_2O + * \quad (5)$$

The suggested ORR mechanism in the presence of OH

$$O_2 + ¤^{OH} + * \rightarrow OOH^* + ¤^O \quad (1)$$
$$OOH^* + (H^+ + e^-) + ¤^O \rightarrow O^* + H_2O + ¤^O \quad (2)$$
$$O^* + (H^+ + e^-) + ¤^O \rightarrow OH^* + ¤^O \quad (3)$$
$$OH^* + (H^+ + e^-) + ¤^O \rightarrow H_2O + * + ¤^O \quad (4)$$

In the presence of the OH-adsorbed group, a revised ORR mechanism is proposed, wherein the formation of OOH* occurs prior to the initiation of the ORR (eq 1-Right) (where ¤ represents a Zr site for -OH adsorption). Indeed, O$_2$ molecules interacting with the -OH group on the hydroxylated Zr surface ($¤^{OH}$) can directly form the OOH* intermediate (eq 1-Right). This eliminates the second step (eq 2-Left) from the conventional four-step ORR mechanism and potentially boosts overall efficiency.

Jia et al[32] experimentally examined the impact of OOH* binding arrangement and Pt surface configuration on ORR activities. Furthermore, Yamamoto et al[31] reported the adsorption energies of ORR intermediates, such as OH*, and OOH*, at defective TiO$_2$ surfaces using density functional theory (DFT). Petersen et al[33] modeled the anion poisoning of a Pt catalyst during ORR and concluded that the predominant factor influencing catalytic activity is the binding of OH*. Despite extensive modeling, additional attention is needed to identify barrier and pathway energies for OOH* creation via O$_2$ and -OH group interactions on catalyst surfaces. As well as a comprehensive understanding of O$_2$ interaction with both defects (V$_O$ or N-doping) and -OH groups simultaneously on the active sites of ZrO$_2$-based catalysts at the atomic level is presently limited.

Based on the above idea, we compare the OOH* intermediate creation on the three hydroxylated catalysts namely t-ZrO$_2$ pristine oxide, under t-ZrO$_{2-x}$ stoichiometry oxide, and t-ZrO$_{2-x}$N$_x$ oxynitride. These catalysts are recognized as promising alternatives to platinum electrocatalysts for future applications. The optimal pathways for OOH* generation on active sites are identified by locating transition states (TS) and determining barrier energy through the CI-NEB method in MD simulations.

To examine the impact of zirconia defects on the migration of hydrogen atom from the OH-adsorbed group to $O_2$ molecule, leading to the forming the OOH* intermediate, we employed the climbing-image nudged elastic band (CI-NEB)[34] method. This approach allow us to calculate the positions and total energies of transition states. By minimizing the energy in both initial and final states, the minimum energy pathway (MEP) of the reaction is determined through the simultaneous minimization of intermediate replicas. A spring force maintains the position of these replicas on the transition path, and the highest energy image ascends the potential energy surface to identify the saddle point representing the activation energy. NEB calculations were conducted using COMB3 potential[35] in LAMMPS[36] software, involving a total of 16 replicas in the analysis with a spring force of 1 kcal/(mol·Å) applied between adjacent images. Further information on the MD simulation setup and NEB calculations is provided in the Supporting Information. Additionally, we utilized DFT calculations to compute the binding energies of the -OH group on both pristine and defective t-$ZrO_2$ surfaces, with detailed explanations available in the Supporting Information.

As already mentioned, surface hydroxylation in t-$ZrO_2$ plays a dual role: potentially poisoning the surface through strong adsorption and aiding the oxygen reduction reaction (ORR) by forming the crucial OOH* intermediate. Before exploring OOH* creation pathways on hydroxylated surfaces, we should first investigate OH-poisoning effects on pristine and defective surfaces.

Zirconium (Zr) sites on a t-$ZrO_2$ serve as active sites for electron transfer. Their surfaces have high binding abilities to -OH. The adsorption energy ($\Delta E$) of an adsorbent (OH group) onto a surface (t-$ZrO_2$) is a key indicator of the degree of surface poisoning, which is calculated by the following equation (eq 1)[37]:

$$\Delta E = E_{\text{(OH-adsorbed- surface)}} - E_{\text{(zirconia slab)}} - E_{\text{(OH)}} \qquad (1)$$

where $E_{\text{(OH-adsorbed- surface)}}$ and $E_{\text{(zirconia slab)}}$ represent the energies of the zirconia surface in the presence and absence of the hydroxyl group, respectively. Also, $E_{\text{(OH)}}$ corresponds to the energy of the isolated hydroxyl group in the gas phase, respectively. The adsorption energy is negative and indicates the strength of the adsorbent-surface interaction. The values of adsorption energy of -OH on pristine t-$ZrO_2$ and two defective t-$ZrO_{2-x}$, and t-$ZrO_{2-x}N_x$ are listed in Table 1.

Table 1. The adsorption energy of -OH on pristine t-$ZrO_2$ and two defective t-$ZrO_{2-x}$, and t-$ZrO_{2-x}N_x$

| | Energy (OH-adsorbed surface) (eV) | Energy (Zirconia slab) (eV) | Energy (Isolated-OH) (eV) | Adsorption Energy (eV) |
|---|---|---|---|---|
| Pristine t-$ZrO_2$ | -627.03 | -616.69 | -6.25 | -4.09 |
| t-$ZrO_{2-x}$ | -620.20 | -606.81 | -6.25 | -7.14 |
| t-$ZrO_{2-x}N_x$ | -615.79 | -627.22 | -6.25 | -5.18 |

The results indicate that the presence of defects affects the affinity of Zr sites toward the –OH group. Adsorption of -OH at the Zr site on both defective surfaces is more stable than the pristine one. In addition, as can be seen in Fig 1, the highest occupied molecular orbital (HOMO) acts as a donor orbital, localizing on $V_o$ and N-dopant sites in t-$ZrO_{2-x}$ and t-$ZrO_{2-x}N_x$ structures, respectively. In contrast, in pristine t-$ZrO_2$, the corresponding orbitals are uniformly distributed across the surface. Notably, introducing defects modifies the surface chemistry of zirconia, making the nearby Zr site more reactive and significantly influencing -OH adsorption strength. It is worth noting that the removal of oxygen ions ($V_o$) creates dangling bonds on the adjacent Zr ions. Therefore, the defect orbital can be highly localized at the two Zr neighbors (Fig 1b). Upon -OH group adsorption on the Zr site adjacent to the vacancy, the oxygen tends to donate electrons to the Zr site. This interaction involves the sharing of electrons and can be considered a form of strong chemical bonding. In contrast, the introduction of a nitrogen dopant induces the formation of a bond between N and O near Zr, with three electrons localized in the π antibonding orbitals (as illustrated in Fig 4S). Moreover, the Zr atom in proximity to the N-dopant displays minimal displacement and contributes fewer electrons to the defect, attributable to the formation of the new N—O bond. Hence, it is understood that the Zr atom in t-$ZrO_{2-x}$ serves as a favorable active site for the adsorption of OH groups, thereby presenting a higher susceptibility to OH poisoning.

Mulliken charges are vital for interpreting adsorption energy, revealing electron distribution among atoms, and insights into charge transfer during adsorption. Mulliken charges for -OH, and Zr ions in the optimized structures of hydroxylated surfaces (pristine t-$ZrO_2$, t-$ZrO_{2-x}$, and t-$ZrO_{2-x}N_x$) are tabulated in Table 2.

Table 2. Mulliken charges on OH, and Zr site for OH adsorption in the optimized structures of OH-poisoned pristine t-$ZrO_2$, t-$ZrO_{2-x}$, and t-$ZrO_{2-x}N_x$ surfaces.

| OH-poisoned surfaces | Mulliken charges on Zr site for OH adsorption | Mulliken charges on O ion of OH | Mulliken charges on H ion of OH | Polarity of Zr-O bond after adsorption on surface |
|---|---|---|---|---|
| Pristine t-$ZrO_2$ | 1.650 | -0.848 | 0.445 | 2.498 |
| t-$ZrO_{2-x}$ | 1.587 | -0.849 | 0.445 | 2.436 |
| t-$ZrO_{2-x}N_x$ | 1.662 | -0.867 | 0.443 | 2.529 |

The results show that the positive charges on Zr ions of pristine t-$ZrO_2$ change in response to different defect types. According to Table 2, the Zr site for -OH adsorption exhibits a Mulliken charge of +1.650 in pristine t-$ZrO_2$, whereas for t-$ZrO_{2-x}$ and t-$ZrO_{2-x}N_x$, the corresponding values are +1.587 and +1.662, respectively. The Mulliken charge of the O ($_{OH}$) ion adsorbed on $Zr_{(t-ZrO2-xNx)}$ is more negative than that of $Zr_{(t-ZrO2-x)}$ and $Zr_{(pristine\ t-ZrO2)}$. We estimate the bond polarity ($\Delta q$) of Zr-O and O-H using the eq 2:

$\Delta q$=Mulliken charge on Atom A−Mulliken charge on Atom B            (2)

where A and B denote Zr and O ions in the Zr-O bond, and once again, A and B represent H and O in the O-H bond, respectively. The bond polarities of Zr-O in both pristine and defective t-$ZrO_2$ are detailed in Table 2. Notably, in $ZrO_{2-x}N_x$, the Zr-O bonds exhibit heightened polarity compared to their counterparts in t-$ZrO_{2-x}$ and pristine t-$ZrO_2$. This observation implies that the separation of H from O adsorbed on Zr(t-$ZrO_{2-x}N_x$) may be more facile in comparison to the other two structures. To validate this, it is imperative to calculate the barrier energy required for the transfer of H from O to an $O_2$ molecule.

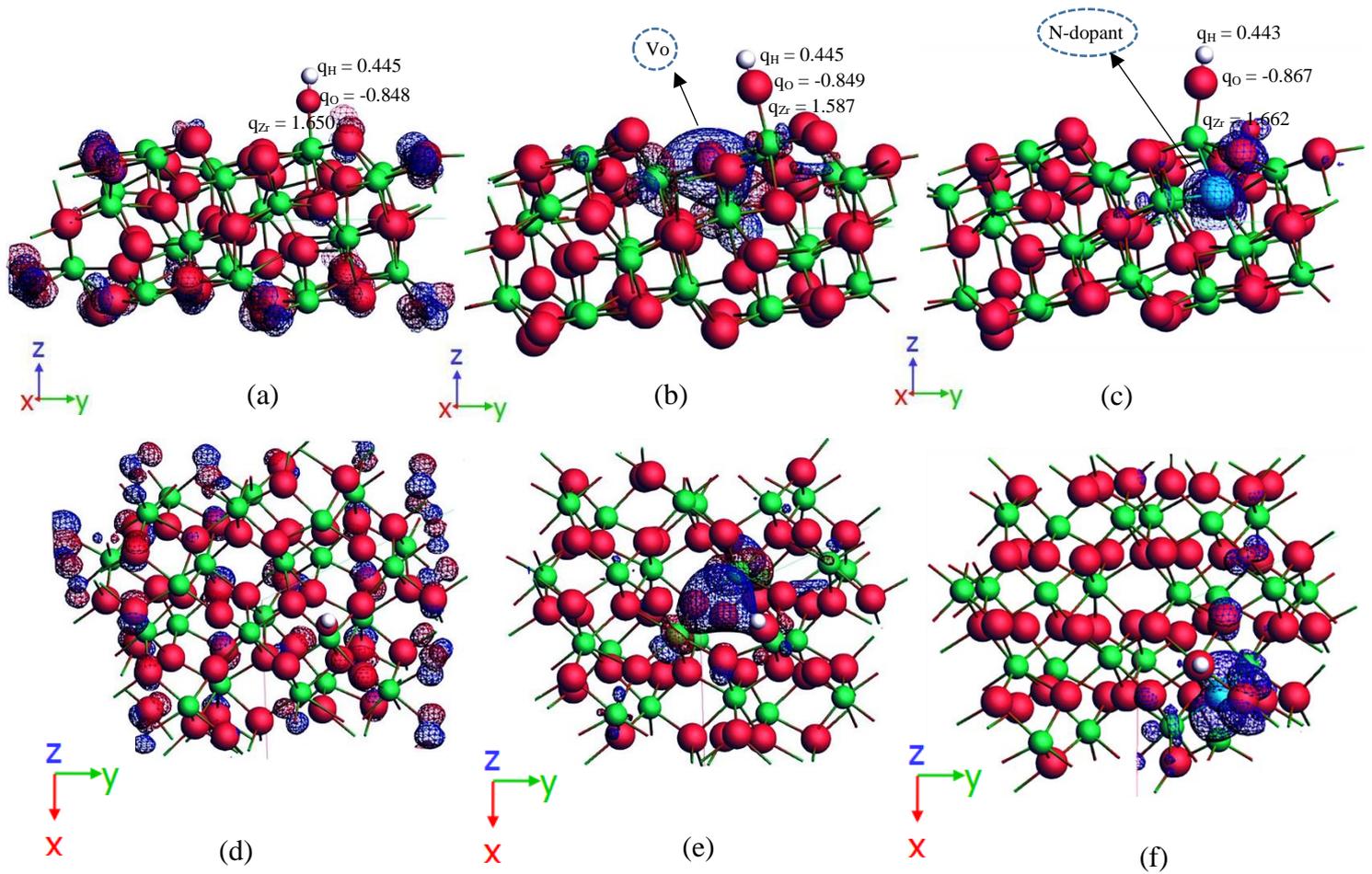

Fig 1. The highest occupied molecular orbital (HOMO) localizing on OH-adsorbed (a) pristine t-ZrO$_{2-x}$ (b) t-ZrO$_{2-x}$, and (c) t-ZrO$_{2-x}$N$_x$ structures, respectively. (d), (e), and (f) illustrate the top view sides of the corresponding orbitals on OH-edsorbed (a) pristine t-ZrO$_{2-x}$ (b) t-ZrO$_{2-x}$, and (c) t-ZrO$_{2-x}$N$_x$ structures, respectively. Note: green, red, blue, and white colors represent Zr, O, N, and H atoms.

To provide insight into the impact of defects on the kinetics of OOH* creation on OH-adsorbed-surfaces, we investigate the activation energy of key reactions. The activation energy barriers are examined by performing NEB calculations on selected reactions' minimum energy path (MEP). To compute MEPs using NEB calculations, it is crucial to bring the O$_2$ molecule close to the OH* group on the surfaces of t-ZrO$_2$-OH, t-ZrO$_{2-x}$-OH, and t-ZrO$_{2-x}$N$_x$-OH. As the O$_2$ molecule progressively approaches the -OH group, the OOH* intermediate forms through the transfer of the H atom from the -OH group to the O$_2$ molecule. Subsequently, the final structures should include the OOH* intermediate adsorbed on Zr atoms, strategically placed in proximity to both the defect and the hydroxylated site. The calculations involve two distinct $O_2$ molecule configurations, including perpendicular, and oblique, relative to OH-adsorbed on zirconia (as shown in Fig 1S).

Initially, we identified transition states between pristine t-$ZrO_2$ (OH) and $O_2$ molecules in two configurations. The final configurations showcased OOH* adsorption on the Zr site, positioned close to the hydroxylated Zr site. Analysis of MEPs reveals that the $O_2$ configurations of initial structures have a minor effect on the reaction pathways of OOH* creation. In pristine t-$ZrO_2$(OH), the barriers for the reaction of adsorbed -OH with $O_2$ molecules in perpendicular, and oblique configurations to form OOH* are 0.16 eV. Correspondingly, the reverse barrier energies for these cases are 0.09 eV. The $O_2$ + OH* → OOH* reaction results in an endothermic process for both $O_2$ configurations. The minimum energy pathways for the explained cases are illustrated in Fig. 2a and Fig. 3.

We conducted more NEB simulations to address the question: "What happens to the barrier for moving a hydrogen ion of -OH on t-$ZrO_{2-x}$ and t-$ZrO_{2-x}N_x$ to $O_2$ molecules with different configurations?" The comparative MEP curves for pristine and two defective OH-adsorbed structures are illustrated in Fig. 2a. In this figure, we exclusively compare the energy pathways of OOH* creation on the three mentioned surfaces in the oblique configuration of $O_2$. Additionally, Fig. 2b illustrates the schematics depicting the initial, transition, and final states of NEB zones for t-$ZrO_{2-x}$-OH. The barrier energies of the reaction $O_2$ + OH* → OOH* on t-$ZrO_2$-OH, t-$ZrO_{2-x}$-OH, and t-$ZrO_{2-x}N_x$-OH have been estimated to be 0.16 eV, 0.26 eV, and 0.11 eV, respectively. In the reverse direction, the associated barrier height is 0.09 eV, 0.25 eV, and 0.1 eV, respectively. Interestingly, the reaction experiences its lowest barrier (0.11 eV) on t-$ZrO_{2-x}N_x$, owing to the formation of a weaker -OH bond with $Zr_{(t-ZrO2-xNx)}$. Furthermore, the anticipated highest barrier energy (0.26 eV) for OOH* creation on the OH-adsorbed-t-$ZrO_{2-x}$ surface suggests strong adsorption of -OH on $Zr_{(t-ZrO2-x)}$ and less interaction between H of- OH and $O_2$.

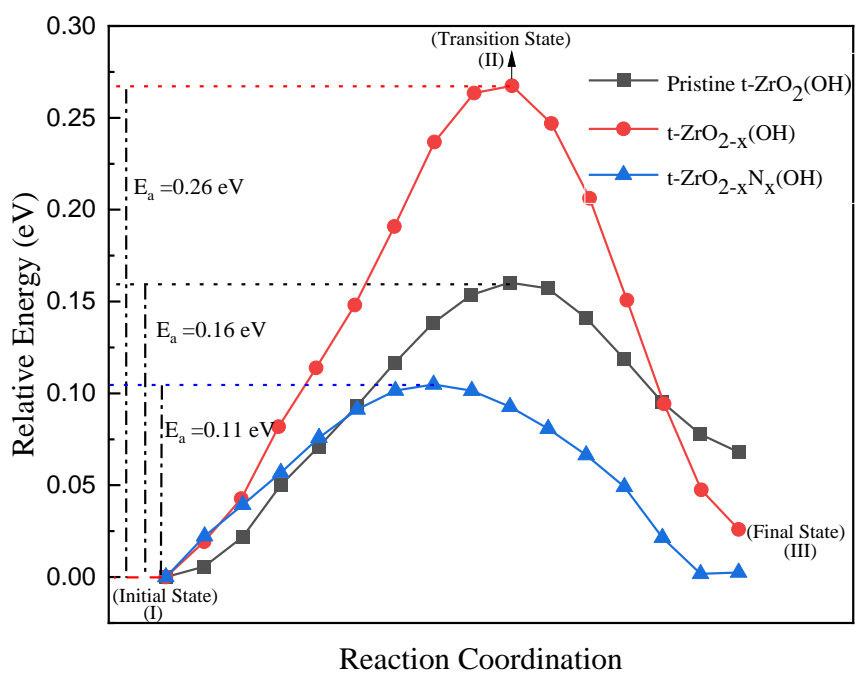

(a)

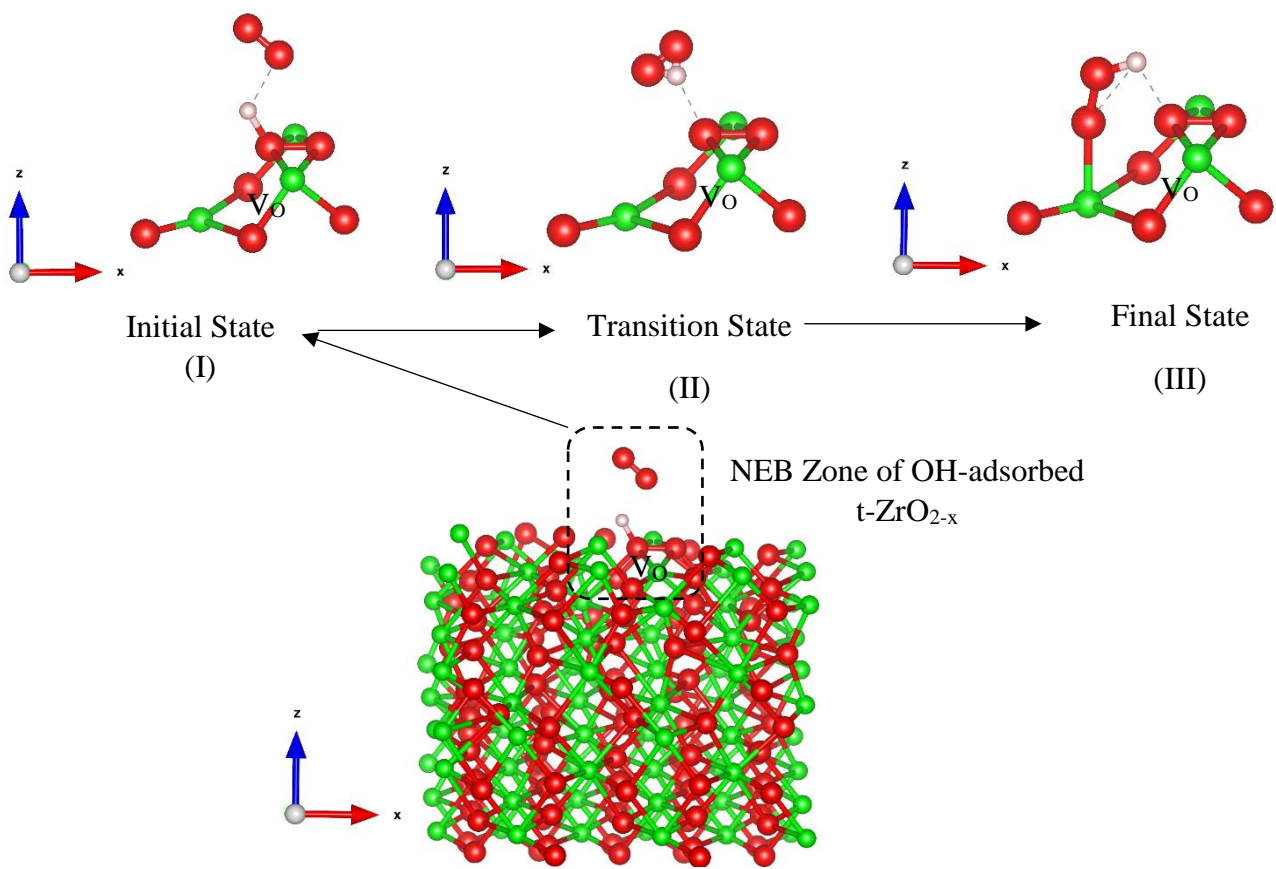

(b)

Fig 2. (a) The minimum energy pathway (MEP) for OOH* creation on OH-adsorbed t-ZrO₂, t-ZrO$_{2-x}$, and t-ZrO$_{2-x}$N$_x$ using the nudged elastic band (NEB) contains the oblique configuration of O$_2$. (b) the schematic representation of the initial, transition, and final states within the NEB zone of OH-adsorbed t-ZrO$_{2-x}$, corresponding to points (I), (II), and (III) in Fig. 2(a).

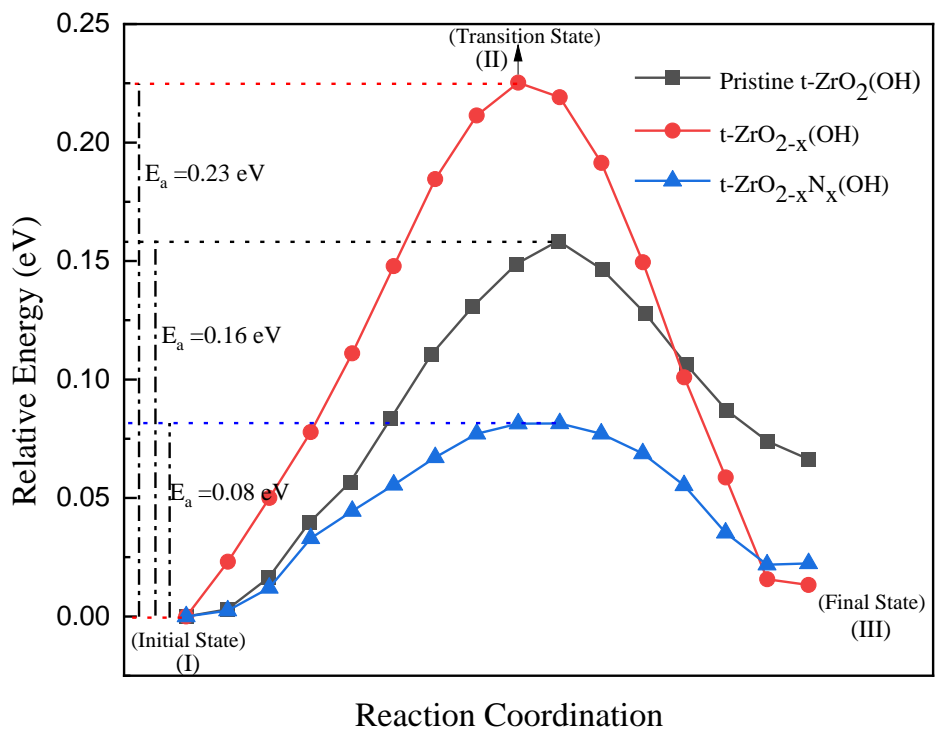

Fig 3. The minimum energy pathway (MEP) for OOH* creation on OH-adsorbed t-ZrO₂, t-ZrO$_{2-x}$, and t-ZrO$_{2-x}$N$_x$ using the nudged elastic band (NEB) contains the perpendicular configuration of O$_2$.

Moreover, the results of MEP curves for OOH* creation on t-ZrO$_{2-x}$-OH, and t-ZrO$_{2-x}$N$_x$-OH under perpendicular configurations of O$_2$ are presented in Fig 3. The results display a decrease in OOH* creation barrier energy from 0.26 to 0.23 eV when changing the O$_2$ configuration from oblique to perpendicular on t-ZrO$_{2-x}$-OH. Additionally, the corresponding value for t-ZrO$_{2-x}$N$_x$(OH) is computed at 0.08 eV when the O$_2$ configuration is perpendicular. Thus, with the perpendicular configuration of the O$_2$ molecule, the creation of OOH* requires lower barrier energies. The values of forward and backward barrier energies of the reaction in all cases are presented in Table 1S. In summary, modeling OH adsorption on zirconia surfaces reveals its dual impact on ORR, serving as both a site blocker and an accelerator for the formation of the crucial OOH* intermediate. This study used DFT calculations to explore the hydroxyl group's adsorption effect on three catalyst surfaces: pristine t-$ZrO_2$, t-$ZrO_{2-x}$, and t-$ZrO_{2-x}N_x$. Subsequently, NEB calculations based on MD simulations were utilized to explore the minimum energy path for OOH* formation on t-$ZrO_2$-OH, t-$ZrO_{2-x}$-OH, and t-$ZrO_{2-x}N_x$-OH surfaces. The results indicate that defects significantly influence the surface's susceptibility to hydroxyl groups. The trend in -OH adsorption energy on pristine and defective surfaces is as follows: t-$ZrO_{2-x}$ > pristine t-$ZrO_2$ > t-$ZrO_{2-x}N_x$. Consequently, t-$ZrO_{2-x}$ is more prone to poisoning with -OH. Additionally, NEB calculations indicated the lowest and highest barrier energy for OOH* creation on OH-adsorbed-t-$ZrO_{2-x}N_x$, and OH-adsorbed-t-$ZrO_{2-x}$, respectively. Furthermore, the perpendicular and oblique $O_2$ configurations minimally impacted OOH* creation barriers in pristine t-$ZrO_2$. The perpendicular $O_2$ configuration on both defective t-$ZrO_2$ results in lower energy levels for OOH* creation. Overall, introducing a nitrogen dopant in t-$ZrO_2$ proves more effective in accelerating ORR activity with reduced -OH adsorption compared to introducing an oxygen vacancy.

## Acknowledgements

This work has been funded by the French National Research Agency (ANR) through the InnOxiCat project ANR-20-CE05–0010.